\documentstyle[12pt]{article}

\def\C{{\rm\kern.24em \vrule width.02em height1.4ex depth-.05ex \kern-.26em
C}}

\newcommand{\beq}{\begin{equation}}
\newcommand{\eeq}{\end{equation}}

\begin{document}
\baselineskip=20pt
\pagestyle{plain}

\title{Liouville Theory: Quantum Geometry \\
of Riemann Surfaces}

\author{
Leon A.~Takhtajan \\
Department of Mathematics\\
SUNY at Stony Brook \\
Stony Brook, NY 1794-3651\\
U.S.A.
}
\maketitle

\begin{center}
{\bf Abstract}
\end{center}
\begin{quote}
Inspired by Polyakov's original formulation \cite{Pol1,Pol2} of quantum
Liouville theory through functional integral, we analyze perturbation expansion
around a classical solution. We show the validity of conformal Ward identities
for puncture operators and prove that their conformal dimension is given by
the classical expression. We also prove that total quantum correction to the
central charge of Liouville theory is given by one-loop contribution, which
is equal to $1$. Applied to the bosonic string, this result ensures the
vanishing of total conformal anomaly along the lines different from those
presented by KPZ \cite{KPZ} and Distler-Kawai \cite{DK}.
\end{quote}

{\bf 1} According to Polyakov \cite{Pol1,Pol2}, basic properties of quantum
Liouville theory can be read from the correlation function
of puncture operators, which is depicted by the following functional integral
\beq \label{FI}
<X>={\cal\int}_{{\cal C}(X)} {\cal D}\phi~e^{-(1/2 \pi h)S(\phi)}.
\eeq
Here $X$ is an $n$-punctured sphere, i.e.~a Riemann sphere $\hat{\C}$ with $n$
removed distinct points, called punctures;
${\cal C}(X)$ is ``domain of integration'', consisting of all smooth conformal
metrics $ds^2=
e^{\phi(w, \bar{w})}|dw|^{2}$ on $X$ satisfying asymptotics
\beq \label{AS}
e^{\phi} \cong \frac{1}{r^{2}_{i} \log^{2}r_{i}}~,\ \ \
i=1, \dots ,n,
\eeq
$r_{i}=|w-w_{i}|,\ i=1, \ldots ,n-1,$ and $r=|w|$ for $i=n$,
near the punctures $w_1, \ldots , w_{n-1}, w_n=\infty$, where $w$ is the
global complex coordinate on $X$; functional $S(\phi)$ is the Liouville action,
and positive $h$ is a coupling constant. Because of asymptotics (\ref{AS}),
naive form of the Liouville action is ill-defined. Its proper definition
is given by the following regularization of the naive action \cite{ZT}
$$
S(\phi)=
\lim_{\epsilon \rightarrow 0} \{  \int_{X_{\epsilon}}(|\phi_{w}|^{2}
+e^{\phi})d^{2}w +2\pi n \log\epsilon +4\pi (n-2)\log|\log\epsilon| \},
$$
where $X_{\epsilon}=
X \setminus  \bigcup^{n-1}_{i=1}\{|w-w_{i}|<\epsilon\}
\bigcup \{|w|>1/\epsilon\}$. Classical equations of motion $\delta S=0$
yield Liouville equation, the equation for complete
conformal metric on $X$ of constant negative curvature $-1$. It has
a unique solution, called Poincar\'{e} metric, and is denoted by
$\phi_{cl}$.

We define functional integral $<X>$ by its perturbation expansion
around the classical solution $\phi= \phi_{cl}$ (cf.~\cite{Zam}, where
$\phi_{cl}$ corresponds to the standard metric of positive constant
curvature on Riemann sphere). It can be obtained from
(\ref{FI}) using the  ``integration measure'' ${\cal D} \phi$
(which is not translation-invariant!), defined by the norm
$$||\delta \phi||^2=\int_{X}|\delta \phi|^{2} e^{\phi} d^{2}w.$$
(cf.~\cite{DK}).
Corresponding propagator $G(w,w^{\prime})$
is given by the Green's function of the operator $2\Delta +1$, where
$$\Delta=e^{-\phi_{cl}} \partial^{2}_{w \bar{w}}$$
is the hyperbolic Laplacian on $X$. Its logarithmic divergence at coincident
points is renormalized in a reparametrization invariant way using the geodesic
distance in the Poincar\'{e} metric
$$G(w,w)=\lim_{w^{\prime} \rightarrow w}\{G(w,w^{\prime})+\frac{1}{2 \pi}(
\log|w-w^{\prime}|^{2} + \phi_{cl}(w))\}.$$

In case of three punctures $w_1, w_2, w_3=\infty,$
it is easy to see that
\beq \label{3}
<X>=\frac{c}{|w_1-w_2|^{1/h}},
\eeq
where $c$ stands for the value of normalized thrice-punctured
sphere with punctures at $0, 1, \infty$. Formula (\ref{3}) supports the
interpretation of $<X>$ as a correlation function of puncture operators
$e^{\phi/2h}$ with conformal dimensions $\Delta=\bar{\Delta}=1/2h$.
Note that the puncture at $\infty$ plays the role of global charge in
accordance with Gauss-Bonnet theorem.
{}From now on we will assume that Riemann surface $X$ is normalized, i.e.~
$X=\C \setminus \{w_1, \ldots , w_{n-3}, 0, 1 \}$.

Conformal invariance of Liouville theory implies that its stress-energy tensor
is traceless. Namely, its $(2,0)$-component $T(\phi)(w)$ is given by
\beq \label{SET}
T(\phi)=\frac{1}{h}(\phi_{ww}-\frac{1}{2}\phi_{w}^{2}),
\eeq
and is conserved on classical equations of motion. It has the following
transformation law under holomorphic change of coordinates
$w=f(\tilde{w})$
$$\tilde{T}(\tilde{w})=T(f(\tilde{w}))(f^{\prime}(\tilde{w}))^{2}+
\frac{1}{h}{\cal S}(f)(\tilde{w}),$$
where ${\cal S}$ stands for the Schwartzian derivative of function $f$
$${\cal S}(f)=\frac{f^{\prime \prime \prime}}{f^{\prime}}-
\frac{3}{2} (\frac{f^{\prime \prime}}{f^{\prime}})^{2}.$$
Geometrically, $T$ has the meaning of a projective connection (times $1/h$).

We emphasize that the modification of Liouville's stress-energy tensor,
i.e.~addition of a total derivative to the ``free-field'' term in
(\ref{SET}), is the crucial feature of the theory. This form of $T(\phi)$ can
be obtained through variation of the generalized Liouville action with
respect to the fiducial metric (see, e.g., \cite{S}).

The expectation value of the stress-energy tensor is defined as follows
$$<T(w)X>={\cal \int}_{{\cal C}(X)} {\cal D}\phi~T(\phi)(w)~
e^{(-1/2\pi h)S(\phi)}.$$
Introducing its normalized value,
$$<<T(w)X>>=<T(w)X>/<X>,$$
one should expect the validity of conformal Ward identities, which we write
in the form
\beq \label{CWI}
<<T(w)X>>=
\sum^{n-1}_{i=1}\frac{\Delta(h)}{(w-w_{i})^{2}} +
\sum_{i=1}^{n-3}(\frac{1}{w-w_{i}}+\frac{w_{i}-1}{w} -\frac{w_{i}}{w-1})
\frac{\partial}{\partial w_{i}}\log <X>,
\eeq
where overall $SL(2, {\bf C})$ symmetry is already fixed.

According to Belavin-Polyakov-Zamolodchikov \cite{BPZ}, conformal Ward
identities (\ref{CWI}) constitute the main axiom of the conformal field
theory in two dimensions and, in principle, should lead to its solution.
Since in our formulation correlation functions $<X>$ and $<<T(w)X>>$ were
already defined by functional integrals, one needs to prove the validity of
(\ref{CWI}) and to calculate conformal dimension $\Delta(h)$.

{\bf 2} It turns out that conformal Ward identities are non-trivial even at
the tree level.
Indeed, one has
$$<<T(w)X>>_{cl}=T(\phi_{cl})(w)=T_{cl}(w)=\frac{1}{h}{\cal S}(J^{-1})(w)$$
$$=\sum_{i=0}^{n-1}\frac{1}
{2h(w-w_{i})^2} + \frac{1}{h}\sum_{i=0}^{n-3}(\frac{1}{w-w_i}+\frac{w_{i}-1}{w}
-\frac{w_{i}}{w-1})c_i$$
and
$$<X>_{cl}=e^{(-1/2\pi h)S_{cl}}.$$
Here ${\cal J}^{-1}$ is the inverse of the uniformization map ${\cal J}$,
which maps the hyperbolic plane $H$ onto the Riemann surface $X$ so that
$X \simeq H/\Gamma$, where $\Gamma$ is the Fuchsian group uniformizing $X$,
$c_{i}$ are the so-called accessory parameters of the Fuchsian uniformization
(introduced by Poincar\'{e}), and $S_{cl}=S(\phi_{cl})$ is the classical
Liouville action. Therefore, conformal Ward identity at the tree level should
imply that
$$\Delta_{cl}=\frac{1}{2h}~~{\rm and}~~c_i=-\frac{1}{2 \pi}\frac{\partial S}
{\partial w_{i}}.$$
The latter formula, conjectured by Polyakov \cite{Pol2}, is quite a
non-trivial statement on accessory parameters (unknown to Poincar\'{e}!),
which was proved in \cite{ZT} (see also \cite{T} for additional discussion)
using Teichm\"{u}ller theory. Note that arguments given above confirm our
definition of the Liouville action and the arrangement of $\pi$'s and $h$'s
elsewhere in the formulas.

Moreover, using the principle that ``central
charge of the theory equals $12$ times the number of Schwartzians in
transformation law of the stress-energy tensor $<<T(w)>>$'' \cite{BPZ},
we see that
$$c_{cl}=\frac{12}{h}=24 \Delta_{cl}.$$
Also note that, according to \cite{Z},
$$S_{cl}=2\pi \log|w_{i}-w_{j}| + O(1),$$
as $w_i \rightarrow w_j,~j=1, \ldots , n-1,$
and
$$S_{cl}=2\pi \log|w| + O(1),$$
as $w \rightarrow \infty$, so that $<X>_{cl}$ indeed has singularities of
a correlation function of primary fields of conformal dimension
$\Delta=\bar{\Delta}=\Delta_{cl}=1/2h$.

{\bf 3} It is also possible to perform one-loop calculations exactly. In doing
so one should use reparametrization invariant regularization of propagator
$G(w,w^{\prime})$ at coincident points, a similar regularization
scheme for its partial
derivatives (using short-distance behavior of Green's functions established
by Hadamard \cite{H}), and the definition of determinant of operator
$(2\Delta + 1)$ through Selberg zeta function of a Riemann surface $X$.
As a result we obtain that \cite{Tak}
$$<<T(w)>>=\frac{1}{h}{\cal S}(J^{-1})(w) + T_{1}(w) + O(h),$$
where
\begin{eqnarray*}
T_{1}(w)=&-&\pi \partial^{2}_{ww^{\prime}}
G(w,w^{\prime})|_{w^{\prime}=w} \\
&-&\pi \int_{X}(G_{ww}(w,w^{\prime})-\phi_{w}(w)
G_{w}(w,w^{\prime}))G(w^{\prime},w^{\prime})e^{\phi(w^{\prime})}d^{2}w^{\prime}.
\end{eqnarray*}

Using regularization scheme described above one can prove the following
properties.

(1) One-loop contribution $T_{1}(w)$ is holomorphic on $X$. It follows from
the definition of Green's function and regularization scheme; the second term
in expression for $T_{1}$, which is the artifact of the modification of the
stress-energy tensor (the first term in (\ref{SET})), is crucial for this
property.

(2) Quantity $T_{1}(w)$ has a transformation law of $1/12$ times projective
connection. This follows from the analysis of the first term
for $T_{1}$, based on the formula
$$\lim_{w^{\prime} \rightarrow w}(\frac{f^{\prime}(w)f^{\prime}(w^{\prime})}
{(f(w)-f(w^{\prime}))^{2}}-\frac{1}{(w-w^{\prime})^{2}})=
\frac{1}{6}{\cal S}(f)(w).$$
The same property is inherent in the free theory and is related with the
second term in (\ref{SET}).

(3) The difference
$$Q_{1}(w)=T_{1}(w)-\frac{h}{12}T_{cl}(w),$$
when lifted to hyperbolic plane $H$ via map ${\cal J}^{-1}$,
defines the holomorphic $\Gamma$-automorphic form of weight $4$
with constant terms $\pi^{2}/12$ at every cusp.

(4) Conformal Ward identities are valid in the one loop approximation with
$$\Delta_{loop}=0.$$

This follows from the property (3), formula
$$\log<X>=-\frac{1}{2 \pi h}S_{cl} -\frac{1}{2}\log\det(2\Delta + 1) +O(h),$$
and explicit expression for the first derivative of Selberg zeta function with
respect to moduli parameters \cite{ZT2}.

Using property (2) and the ``exchange rate''
central charge/$12$ times Schwartzians, we obtain
$$c_{loop}=12 \times \frac{1}{12} =1,$$
and, therefore,
$$c_{Liouv}=c_{cl}+c_{loop}=\frac{12}{h}+1,~~\Delta=\Delta_{cl}=
\frac{1}{2h}.$$

{\bf 4} It is remarkable that simple expression
$$c_{Liouv}=1+\frac{12}{h}$$
for the central charge of the theory is exact. Indeed, one can
prove this theorem analyzing every term in perturbation expansion of
$<<T(w)X>>$. Although it looks like almost an impossible task, we observe that
Schwartzians in higher loops can only appear through contribution of the
second term in definition (\ref{SET}) of the stress-energy tensor. This is the
term which corresponds to the free theory and it is easy to keep its track in
all orders of perturbation theory, which completes the proof.

It should be noted that this formula for the central charge was conjectured in
\cite{ST} on quantum group symmetry of the Liouville model.

Our results look different from traditional ones (cf.~\cite{KPZ,DK,DH}).
We note that, contrary to the approach in \cite{DK}, we do not make any a
priori
assumptions about the change of the ``integration measure'' in functional
integral. Instead, we systematically use properties of the classical
solution---
Poincar\'{e} metric---and unambiguously define our main
quantities---correlation
functions $<X>$ and $<<T(w)X>>$---through perturbation expansion. However,
we do not know yet how to treat these basic objects of ``quantum
geometry of Riemann surfaces'' non-perturbatively. We also observe that whereas
in approaches of KPZ and Distler-Kawai, one always has $c < 1$, in our
approach, it is always $c >1 $. This might suggest that we are actually dealing
with two different phases of $2-D$ quantum gravity, separated by the
``$c=1$ barrier'' (cf.~discussion in \cite{Pol3}).

Details of calculations, generalization to the higher genus case, and precise
formulation of auxiliary mathematical results we have used, will be presented
elsewhere.

Here, as a final speculation, we point out that if
$$\frac{1}{2\pi h}=\frac{25-D}{24 \pi},$$
as one should expect from the bosonic string in $D$ dimensions, then
$$c_{Liouv}=25-D+1=26-D.$$
This is the expression one really needs: it cancels the contribution from
string modes and ghosts and ensures the vanishing of total conformal anomaly
for any $D$.

{\bf Acknowledgments.} I appreciate useful discussions with S.~Shatashvili.
This research was supported in part by the NSF grant DMS-92-04092.

\end{document}